\documentclass[twocolumn]{aastex631}

\usepackage{longtable}
\usepackage{graphicx}
\usepackage{amsmath}	

\newcommand{\siggas}{$\Sigma_{\rm gas}$}

\newcommand{\sightwo}{$\Sigma_{\rm H_{2}}$}
\newcommand{\sighone}{$\Sigma_{\rm HI}$}
\newcommand{\sigsfr}{$\Sigma_{\rm SFR}$}
\newcommand{\sigstar}{$\Sigma_{\ast}$}

\newcommand{\re}{$R_{\rm e}$}
\newcommand{\mstar}{$M_{\ast}$}
\newcommand{\msun}{$M_{\odot}$}
\newcommand{\Av}{$A_{\rm V}$}
\newcommand{\OIII}{\hbox{[O\,{\sc iii}]}}
\newcommand{\OII}{\hbox{[O\,{\sc ii}]}}
\newcommand{\NII}{\hbox{[N\,{\sc ii}]}}
\newcommand{\SII}{\hbox{[S\,{\sc ii}]}}
\newcommand{\Ha}{H${\alpha}$}

\newcommand{\HI}{\hbox{H\,{\sc i}}}
\newcommand{\HII}{\hbox{H\,{\sc ii}}}
\newcommand{\dlre}{$\Delta\,$log$\,R_{\rm e}$}
\newcommand{\dlsfr}{$\Delta\,$log$\,$SFR}
\newcommand{\dssfr}{$\Delta\,$sSFR}
\newcommand{\loh}{$\rm 12+log(O/H)$}
\newcommand{\dindex}{D$_{n}$(4000)}
\newcommand{\ewhd}{EW(H$\delta_A$)}
\newcommand{\ewha}{EW(H$\alpha$)}

\defcitealias{Dopita-16}{DOP16}
\defcitealias{Pilyugin-Grebel-16}{PG16}

\shorttitle{Resolved properties across the mass-size plane}
\shortauthors{Lin et al.}

\begin{document}

\title{Radial Profiles of \sigstar, \sigsfr, Gas Metallicity and Their Correlations Across the Galactic Mass-Size Plane}

\author{Lin Lin}
\affiliation{Shanghai Astronomical Observatory, 80 Nandan Road, Shanghai, China}
\email{linlin@shao.ac.cn}

\author{Shiyin Shen}
\affiliation{Shanghai Astronomical Observatory, 80 Nandan Road, Shanghai, China}

\author{Hassen M. Yesuf}
\affiliation{Shanghai Astronomical Observatory, 80 Nandan Road, Shanghai, China}

\author{Ye-Wei Mao}
\affiliation{Center for Astrophysics, GuangZhou University, GuangZhou 510006, People's Republic of China}
\affiliation{Department of Astronomy, School of Physics and Materials Science, Guangzhou University, GuangZhou 510006, People's Republic of China}

\author{Lei Hao}
\affiliation{Shanghai Astronomical Observatory, 80 Nandan Road, Shanghai, China}


\begin{abstract}

We analyzed the global and resolved properties of approximately 1,240 nearby star-forming galaxies from the MaNGA survey, comparing compact and extended galaxies -- those with smaller and larger radii (\re), respectively -- at a fixed stellar mass (\mstar). Compact galaxies typically exhibit lower \HI\ gas fractions, higher dust extinction, higher metallicity, greater mass concentration, and lower angular momentum on a global scale. Radial profiles of stellar mass surface density (\sigstar) and star formation rate surface density (\sigsfr), as functions of the effective radius ($R/$\re), reveal that compact galaxies display steeper gradients and higher values, resulting in elevated specific star formation rates (sSFR) in their inner regions compared to their outskirts. At a given \sigstar, compact galaxies have higher sSFR than extended galaxies, particularly in low-mass galaxies (log(\mstar/$M_{\odot}$)$\,\leq\,$10$^{10}$). Additionally, their metallicity profiles differ significantly: extended galaxies have steeper metallicity gradients, while compact galaxies exhibit flatter slopes and higher metallicity at a given $R/$\re. After accounting for the dependence of metallicity on \mstar\ and \sigstar, no further correlation with SFR is observed. The combination of higher sSFR and potentially higher star formation efficiency in compact galaxies suggests that their central gas is being rapidly consumed, leading to older stellar populations, as indicated by \dindex\ and \ewhd, and resulting in faster central growth. Our results reveal that radial SFR profiles cannot be fully determined by \mstar\ and \sigstar\ alone; other factors, such as galaxy size or angular momentum, must be considered to fully understand the observed trends.

\end{abstract}



\section{Introduction} \label{sec:intro}

Galaxies are complex and evolved ecosystems with many parameters to describe them, such as stellar mass (\mstar), star formation rate (SFR), and gas-phase metallicity (Z or 12+log(O/H)). Stellar mass represents the bulk of the mass in galaxies and serves as a record of a galaxy's past star formation history. SFR provides a measurement of the rate at which galaxies are forming new stars, which is related with gas accretion, gas cooling, the formation and collapse of molecular clouds, and stellar feedback. Moreover, the gas-phase metallicity reflects the cumulative enrichment of heavy elements by past stellar generations, influences star formation efficiency and relates to gas inflow and outflow processes as well. Radial profiles of these parameters provide insights into galaxy formation and evolution.

Many scaling relations are built up from observation to understand the underlying physical processes. The star formation main sequence (SFMS) refers to a tight correlation between the star formation rate (SFR) and the stellar mass (\mstar) of galaxies \citep{Brinchmann-04, Daddi-07, Noeske-07, Speagle-14, Renzini-Peng-15}. Similarly, the mass-metallicity relation (MZR) refers to the empirical correlation between the stellar mass and the gas-phase metallicity of galaxies \citep{Lequeux-79, Tremonti-04, Maiolino-08, Zahid-14}. 
Further studies proposed the fundamental metallicity relation (FMR) as a refinement of the mass-metallicity relation (MZR) to account for the observed scatter in metallicity at fixed stellar mass \citep{Ellison-08a, Mannucci-10, Lara-Lopez-10}. It suggests that galaxies with the same stellar mass but different star formation rates may have different gas-phase metallicities, with higher star formation rates associated with lower metallicities at fixed stellar mass.

As the development of integral-field spectroscopy (IFS), studies found that these relations still hold on small scales for star formation regions, i.e. the resolved star formation main sequence (rSFMS) -- the surface densities of SFR (\sigsfr) are well correlated with the surface densities of stellar mass (\sigstar) \citep{Rosales-Ortega-12, Wuyts-13, Cano-Diaz-16, Nelson-16, Hsieh-17}, the resolved mass-metallicity relation (rMZR) -- the variations of metallicity within galaxies are correlated with the surface densities of stellar mass (\sigstar) \citep{Rosales-Ortega-12, Sanchez-13, Barrera-Ballesteros-16} and the resolved fundamental metallicity relation (rFMR) which describes the relationship among \sigsfr, \sigstar, and gas metallicity \citep{Maiolino-Mannucci-19, Sanchez-Menguiano-19, Sanchez_Almeida-Sanchez-Menguiano-19}.

In spite of lots of studies on these scaling relations, the role of global and local properties in shaping the evolution of galaxy is still unknown. Some studies argue that the global relations are just the integrated versions of the local ones, suggesting that the evolution of galaxies is ruled by local properties  \citep{Sanchez-20, Sanchez-21a}. \citet{Cano-Diaz-19} demonstrated that the global SFMS is a consequence of the local one and the local star-formation regions follow an more universal process of SF. \citet{Barrera-Ballesteros-16} studied the local MZR and found it can reproduce both the global MZR and radial metallicity gradients. \citet{Sanchez_Almeida-Sanchez-Menguiano-19} shown that the global FMR can be derived analytically from the local anti-correlation between \sigsfr\ and Z.  

However, others argue that the global relations are not simply a sum of their resolved counterparts. For instance, several studies have found that the \sigsfr\ can be linked to the global kinematics of the galaxy \citep{Kennicutt-98b, Tan-00, Aouad-20}, despite that star formation is believed to be happened on small scale and be associated with \siggas\ or \sigstar. \citet{Pezzulli-21} found that the evolution of the mass-size relation cannot be simply reproduced from the resolved SFMS, raising questions about whether the resolved main sequence is fundamentally local. As for the metallicity relations, \citet{Baker-23} determined the relative importance of local and global properties and found global properties like gravitational potential, large-scale winds and the global mixing of metals, contribute to the local metallicity. Similar result has also been reported in \citet{Gao-18}.


%
In this work, we aim to analysis the radial profiles of \sigstar, \sigsfr, \loh, and revisit their scaling relations in different sizes of galaxies at fixed stellar mass. We consider that comparing the resolved properties between compact and extended galaxies at fixed stellar mass could help us decouple the connection between global and local properties to some extent. First, stellar mass serves as a key parameter in various scaling relations that link different properties of galaxies. By fixing the global stellar mass, we can construct a subset with similar characteristics in many aspects, such as the global SFR, halo mass, etc. Second, since surface densities are proportional to $r^{-2}$, we would achieve a wide range of surface densities. Considering the observed mass-size relation for SF galaxies, the scatter in radius at fixed stellar mass is 0.3 dex \citep{Shen-03, van_der_Wel-14}, leading to a difference of 1.2 dex in surface densities. 

Comparing two local star formation regions with similar densities -- one representing an outer region in massive and extended galaxies, and the other an inner region in less massive and compact galaxies -- they probably follow the same scaling relations if local properties are indeed the primary factor. If not, it implies that factors such as galaxy size \citep{Mo-98} and the various dynamical processes in the evolutionary history \citep{DeFelippis-17} should be taken into account in the star formation process.

Using the Mapping Nearby Galaxies at Apache Point Observatory \citep[MaNGA; ][]{Bundy-15} dataset, our goal is to present a homogeneous study for star formation galaxies that spans a range of masses and sizes. Instead of studying the galaxy-to-galaxy variations, our work focus on the general trends (median radial profiles) as a function of mass and size. This approach enables a better understanding of the properties within compact galaxies, which may be overlooked in previous studies if all spaxels are aggregated indiscriminately.

The paper is organized as follows. Section \ref{sec:data} describe our data and sample selection, along with our measurements of the median radial profiles in each mass-size bin. We present the main results in Section \ref{sec:comparison} and \ref{sec:scaling_relations}. Discussion and summary are provided in Section \ref{sec:discuss} and \ref{sec:summary}, respectively. Throughout this paper, we adopt a Salpeter initial mass function (IMF) and a cosmology with $H_{0}$ = 70 km s$^{-1}$ Mpc$^{-1}$, $\Omega_{\rm M}$ = 0.3, and $\Omega_{\Lambda}$ = 0.7.

\section{Sample and Data Reduction} \label{sec:data}

\subsection{The MaNGA Survey}

In this work, we used the publicly available data cubes of the MaNGA survey, which is an integral-field unit (IFU) spectroscopic survey that includes more than $\sim$10,000 galaxies \citep{Bundy-15, Drory-15, Yan-16b, Wake-17}. For each galaxy, the released data cube has a spatial coverage out to 1.5 -- 2.5 \re\ (for Primary and Secondary sample respectively) and a spectral resolution of $R\sim$ 2000 covering wavelengths from 3600 to 10,000 \AA. The typical $r-$band signal-to-noise ratio is about 4-8 (per \AA\ per 2\arcsec-fiber) at effective radius of 1-2.

The advanced data products, including emission line flux and kinematics measurements, are provided by MaNGA Data Analysis Pipeline (DAP) \citep{Westfall-19, Belfiore-19}. Specifically, we use the ``hybrid'' binning products (HYB10-MILESHC-MASTARHC2) in this work, for which the stellar spectrum and stellar kinematics are obtained by applying pPXF code to binned spectra (S/N$\sim$10), while the emission lines are fitted for each spaxel with Gaussian profiles \citep{Belfiore-19}. Bad spaxels that might be contaminated by foreground stars or have critical failures during data reduction are removed by the bitmask adopted in the DAP.

\subsection{Sample Selection}

\begin{figure}
	\begin{center}
		\includegraphics[width=0.45\textwidth]{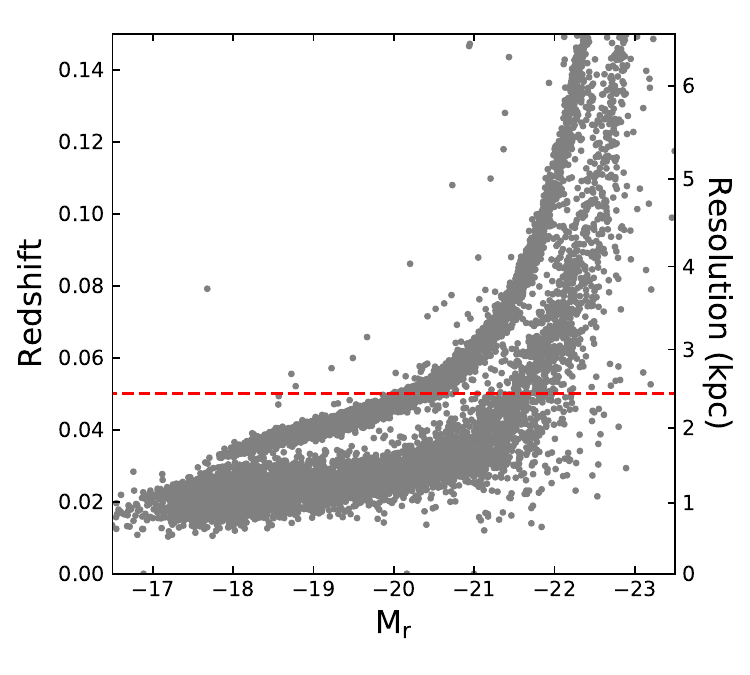}
	\end{center}
	\caption{Distribution of MaNGA galaxies on the diagram of redshift vs. $r$-band absolute magnitude. The spatial resolution, corresponding to the physical scale of a typical PSF FWHM of 2\farcs5, is also shown along the ordinate. The galaxies populate two branches, corresponding to the MaNGA Primary and Secondary samples. To ensure sufficient spatial resolution, we limit our sample galaxies at redshift $z<$ 0.05 with axial ratios b/a $>$ 0.5.}
	\label{fig:resolution}    
\end{figure}

\begin{figure}
	\begin{center}
		\includegraphics[width=0.45\textwidth]{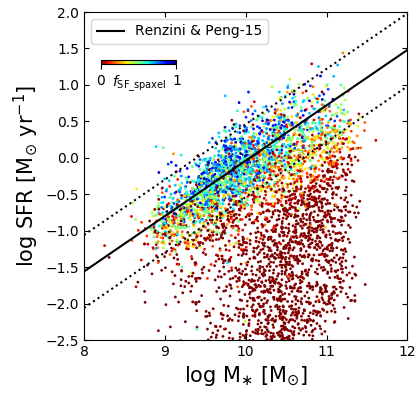}
	\end{center}
	\caption{Selection of star formation galaxies. Star formation galaxies are selected by requiring their locations on the star formation main sequence with a width of 0.5 dex and the fractions of the star formation spaxels are larger than 50\%.}
	\label{fig:SFMS_cut}    
\end{figure}

\begin{figure}
	\begin{center}
		\includegraphics[width=0.45\textwidth]{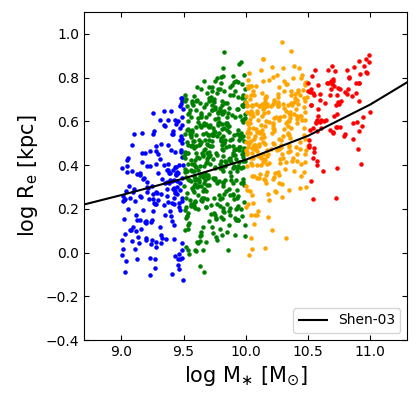}
	\end{center}
	\caption{Sample distribution in the mass-size plane. The selected star-forming galaxies are separated into different mass-size bins. The solid line is the mass-size relation for late-type galaxies in SDSS \citep{Shen-03}. }
	\label{fig:mass_size_bins}    
\end{figure}

\begin{figure*}
	\begin{center}
		\includegraphics[width=\textwidth]{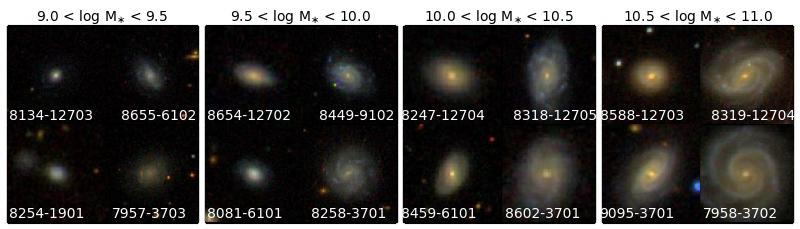}
	\end{center}
	\caption{Color images of a random sample of MaNGA star-forming galaxies, illustrating the difference between compact and extended galaxies. Images are composed from $gri$ bands. Within each mass bin, two compact galaxies are displayed on the left, while two extended galaxies are shown on the right. All images are scaled to cover 30 kpc on each side. }
	\label{fig:jpeg_example}    
\end{figure*}

Our aim is to study the resolved profiles within star formation galaxies, for which we select galaxies based on the following criteria.

To ensure galaxies with good spatial resolution, we restricted our sample to galaxies with relatively low redshifts ($z$ $<$ 0.05) and nearly face-on galaxies ($b/a$ $>$ 0.5), where the semi-major ($a$) and semi-minor ($b$) axes are taken from the NSA catalog and are measured at the 25 mag arcsec$^{-2}$ isophote in the $r$-band. As can be seen from Figure~\ref{fig:resolution}, this redshift cut results our sample with a median spatial resolution of $\sim$1.5 kpc. This cut reduces the sample to 4482 galaxies.

We select star formation galaxies by their locations in the star formation main sequence. Stellar mass and SFR are taken from GALEX-SDSS-WISE legacy catalog based on UV-optical SED fitting \citep[][X2-version]{Salim-16, Salim-18}. The SFMS ridgeline from \cite{Renzini-Peng-15} was adopted to calculate the SFR residuals at fixed stellar mass ($\Delta$log SFR). We select star formation galaxies by requiring they have $| \Delta\,$log SFR$|$ $<$ 0.5 dex (2188 galaxies). In addition, we also require the fraction of the SF spaxels are larger than 50\% following \cite{Hsieh-17}. In specifically, We adopt the BPT diagram (\OIII/H$\beta$ versus \NII/H$\alpha$ version) to select \HII\ spaxels according to the curve in \cite{Kauffmann-03}. We calculate the ``SF fraction'' ($f_{\rm SF}$) as the ratio between the number of \HII\ spaxels and the number of spaxels within 1.5 \re. All spaxels should have reliable measurements (SNR $>$ 3) for the four emission lines in the BPT diagram. Figure \ref{fig:SFMS_cut} shown the SFR-\mstar\ relation and colored by the $f_{\rm SF}$. There are 1270 star formation galaxies left in our sample.

Additionally, previous studies have suggested that galaxy interactions/disruptions can enhance central star formation and lower the metallicity \citep{Ellison-08b, Kewley-10}. However, more recent work indicates that these effects may be less common, with the changes in SFR and metallicity likely occurring only in the late stages of the merger process \citep{Scudder-12, Pan-18}. To assess the environmental impact on our sample, we first cross-matched the selected galaxies with MaNGA pair catalog, which identifies pairs by a projected separation $<$ 200 kpc and a line-of-sight velocity difference $<$ 500 km s$^{-1}$ \citep{Feng-20}. We also matched our sample with MaNGA Visual Morphologies catalog, which identifies visible tidal features using SDSS and DESI images \citep{Vazquez-Mata-22}. This process yielded 364 candidates with either close companions or tidal features.  After visually inspecting these galaxies using their SDSS images, we found that most have regular optical disks. Consequently, we excluded only 28 galaxies with clear signs of interaction or disruption (lack of regular disk structures). We have checked that excluding all 364 candidates would not change our main results.

At the end, we obtain 1242 galaxies. Figure \ref{fig:mass_size_bins} shows their distributions in the mass-size plane. We characterized galaxy size using the semi-major axis in $z-$band images, with the radius values obtained from the NYU Value-Added Galaxy Catalog \citep{Blanton-05}. Different colors represent galaxies with different stellar masses. Within each mass bin, compact and extended galaxies are separated according to the mass-size relation for late-type galaxies in SDSS \citep{Shen-03}. 
Figure \ref{fig:jpeg_example} shows the color images of both compact and extended galaxies. In each mass bin, two compact galaxies are shown on the left with two extended ones on the right.

\subsection{Measurements of Stellar Mass, SFR, and Gas Metallicity}

\subsubsection{Stellar Mass Surface Density, \sigstar}

We adopt the two-dimensional maps of stellar mass surface density (\sigstar) provided by \citep{Lu-23}, in which they performed stellar population synthesis based on FSPS model \citep{Conroy-09, Conroy-Gunn-10}, including 43 age and 9 metallicity grids. The Salpeter IMF was adopted. The datacubes are Voronoi binned to SNR$\sim$30 before spectral fitting to obtain reliable stellar population maps. 

We note that some profiles in outer region become flat due to the large bins. To avoid the artificial flat data points, we only use the measurements with $g$-band SNR $\ge$ 5.

\subsubsection{Star Formation Rate Surface Density, \sigsfr}

We estimate the SFR density from the dust corrected H$\alpha$ luminosity using the calibration given in \citet{Kennicutt-98a}: SFR = 7.9$\times 10^{-42} L$(H$\alpha$), with the assumptions of Salpeter IMF, an initial H${\alpha}$/H${\beta}$ ratio of 2.86, and the Galactic extinction curve with $k(V)$ = 3.1.

Only star-forming spaxels are taken into account in calculating the profiles of \sigsfr. As previous works suggested that LIERS also contribute H$\alpha$ emission, we select spaxels that are classified as star-forming using the \NII/H$\alpha$ versus \OIII/H$\beta$ BPT diagram ($y = 0.61 / (x - 0.05) + 1.3$ \citep{Kauffmann-03}, where $y$ = log(\OIII/H${\beta}$) and $x$ = log(\NII/H{$\alpha$})) and the SNR $>$ 3 for the four emission lines.  Regions do not satisfied the above criteria are considered as zero star formation.

\subsubsection{Gas-phase Metallicity, 12+log(O/H)}

We determine the gas-phase metallicity using the {\tt R} calibration described in \citet[][hereafter \citetalias{Pilyugin-Grebel-16}]{Pilyugin-Grebel-16} and the {\tt N2S2Ha} calibrator of \citet[][hereafter \citetalias{Dopita-16}]{Dopita-16}. Both calibrations involve the \NII/\Ha\ term, which is considered to be less sensitive to the ionization parameter \citep{Zhang-17, Hwang-19}.

The {\tt R} calibration in \citetalias{Pilyugin-Grebel-16} relies on the following three diagnostic line ratios:
\begin{equation}
\begin{split}
    & \rm {\tt R2} = \OII\lambda\lambda3727,3729/H{\beta}, \\  
    & \rm {\tt R3} = \OIII\lambda\lambda4959,5007/H{\beta}, \\
    & \rm {\tt N2} = \NII\lambda\lambda6548,6584/H{\beta}. 
\end{split}
\end{equation}

For star formation regions with log\,{\tt N2} $\ge -0.6$ (the upper branch), the metallicity is calculated as 
\begin{equation}
\begin{split}
    \rm 12+log(O/H) = & \, \rm 8.589+0.022\,log\,({\tt R3}/{\tt R2}) \\
            & \rm +0.399\,log\,{\tt N2} + (-0.137 \\
            & \rm +0.164\,log\,({\tt R3}/{\tt R2}) \\ 
            & \rm +0.589\,log\,{\tt N2}) \times log\,{\tt R2}, 
\end{split}
\end{equation}
and the metallicity for the lower branch (log\,{\tt N2} $< -0.6$) is calculated as
\begin{equation}
\begin{split}
    \rm 12+log(O/H) = & \, \rm 7.932+0.944\,log\,({\tt R3}/{\tt R2}) \\
            & \rm +0.695\,log\,{\tt N2} + (0.970 \\
            & \rm -0.291\,log\,({\tt R3}/{\tt R2}) \\ 
            & \rm - 0.019\,log\,{\tt N2}) \times log\,{\tt R2}.
\end{split}
\end{equation}

Besides, the {\tt N2S2Ha} calibrator proposed by \citetalias{Dopita-16} is expressed as
\begin{equation}
\begin{split}
    \rm {\tt N2S2Ha} = & \, \rm log(\NII\lambda6584/\SII\lambda\lambda6717,6731) \\
                & \rm +0.264\,log(\NII\lambda6584/H{\alpha}).
\end{split}
\end{equation}
Then the empirical relation between metallicity and {\tt N2S2Ha} is
\begin{equation}
    \rm 12+log(O/H) = 8.77+{\tt N2S2Ha}+0.45\,({\tt N2S2Ha}+0.3)^5.
\end{equation}

Similar to SFR calculations, only star-forming spaxels are taken into account when computing the 12+log(O/H) values.

\subsubsection{Radial Profiles in Each Mass-Size Bin}

Based on the two-dimensional maps within individual galaxies, we first calculate the one-dimensional radial profiles of different parameters. The position angle of the major axis (PA) and the major-to-minor axis ratio ($b/a$) from the NSA catalog were used to construct the deprojected one-dimensional radial profiles for each galaxy. As we separated galaxies into different mass-size bins, the median profiles are then derived by calculating the median values of individual profiles along radius with a sampling of 0.2 $R/R_{\rm e}$ in each mass-size bin. Errors are estimated by calculating the sample standard deviation ($\sigma$) and dividing by $\sqrt{N}$, where $N$ is the number of profiles at each radius. 

We primarily focus on the \sigstar, \sigsfr, and \loh\ median profiles in the main text, additional median profiles such as \Av, \dindex, \ewhd, \ewha\ and individual profiles for each mass-size bin are presented in Appendix \ref{app:profiles}. Different colors indicate different stellar masses, with solid and dashed lines representing extended and compact galaxies, respectively.

\begin{figure*}
	\begin{center}
            \includegraphics[width=0.8\textwidth]{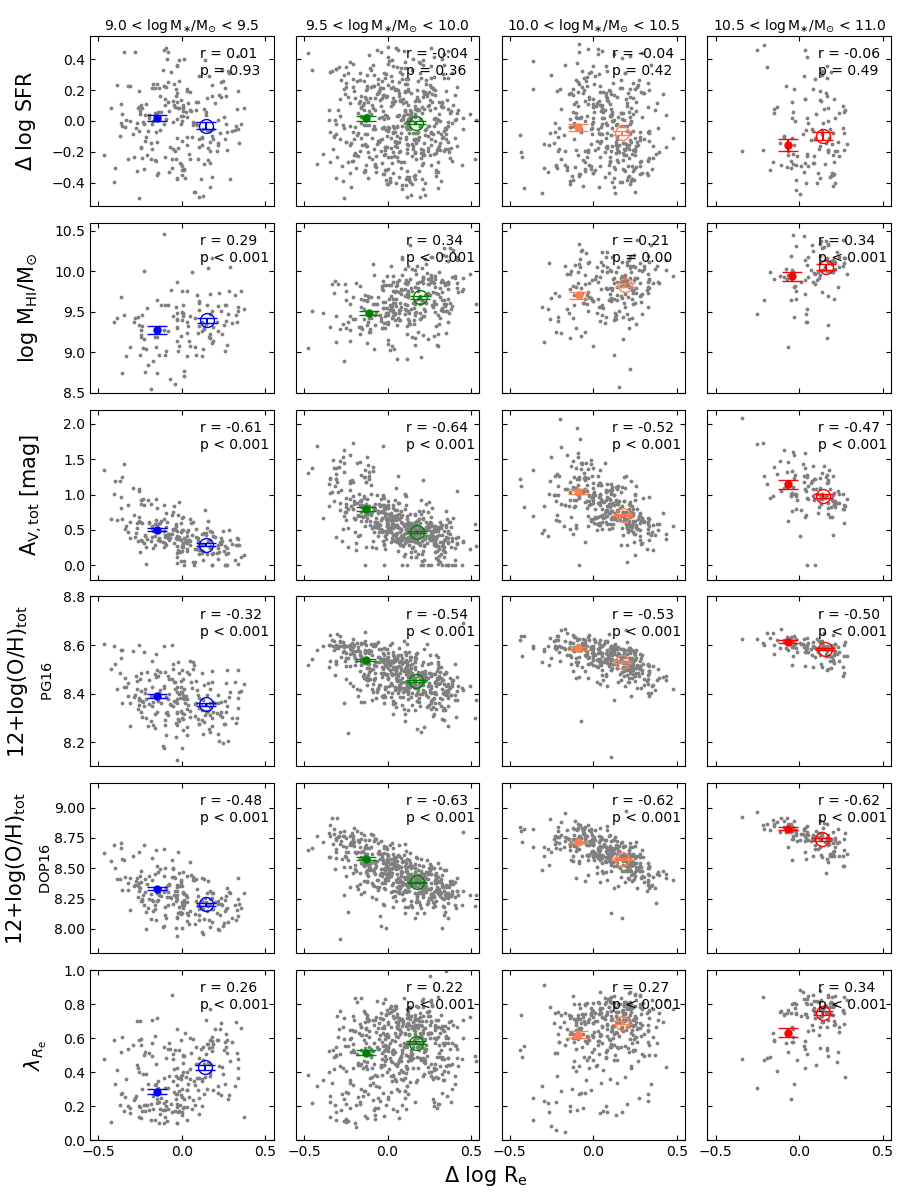}
	\end{center}
	\caption{Distribution of star formation residual $\Delta$ log SFR, \HI\ mass log M$_{\rm HI}$, dust extinction \Av, \loh, stellar spin $\lambda_{R_{\rm e}}$ as a function of radius residual $\Delta$ log \re.  $\Delta$ log SFR and $\Delta$ log \re\ represent the residuals from the star-forming main-sequence and the mass-size relation, respectively. Grey dots are the star-forming galaxies selected in our sample. Larger symbols show the median values for compact and extended galaxies, with different colors indicating different mass bins. Error bars indicate their median errors. The Pearson correlation coefficient r and the p-values are listed in the top-right corner of each panel.}
	\label{fig:global}    
\end{figure*}

\begin{figure*}
    \centering
    \includegraphics[width=0.8\textwidth]{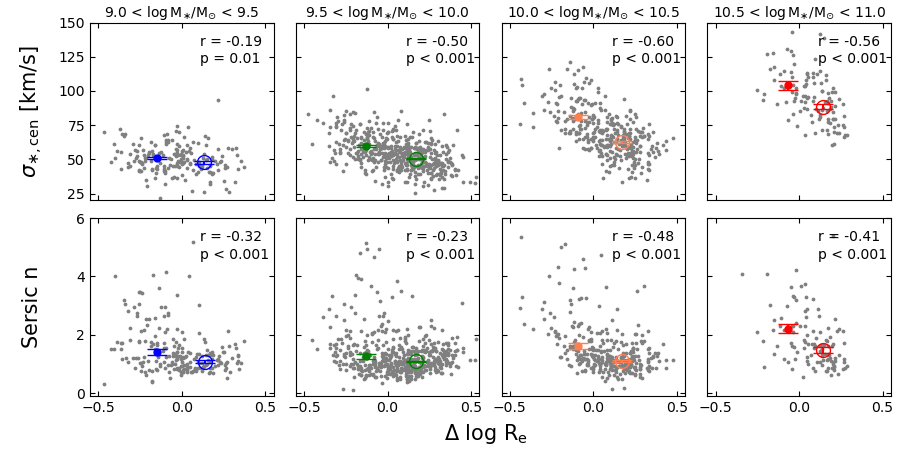}
    \caption{Same as Figure \ref{fig:global}, for central stellar velocity dispersion $\sigma_{\ast, \rm cen}$ and Sersic index $n$.}
    \label{fig:global2}   
\end{figure*}

\begin{figure*}
	\begin{center}
            \includegraphics[width=0.8\textwidth]{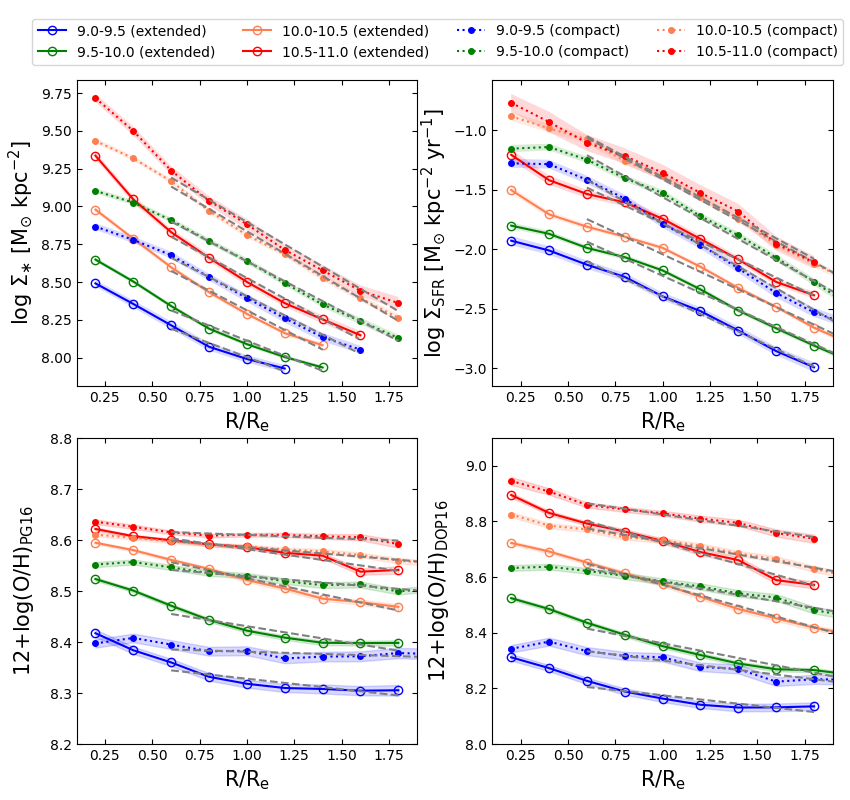}
	\end{center}
	\caption{Radial profiles of \sigsfr, \sigstar, \loh$_{\rm PG16}$ and \loh$_{\rm DOP16}$ in different mass-size bins. The color of the lines represents different stellar mass bins. In each mass bin, dotted lines with filled dots are the median profile from compact galaxies, while solid lines with empty circles are the median profiles from extended galaxies. The shaded region represents the error on the median profile. The grey dashed line is the linear fit to the median profile (in radial range $>0.5$\re) in each mass-size bin.}
	\label{fig:radprofiles}    
\end{figure*}

\begin{figure}
	\begin{center}
            \includegraphics[width=0.48\textwidth]{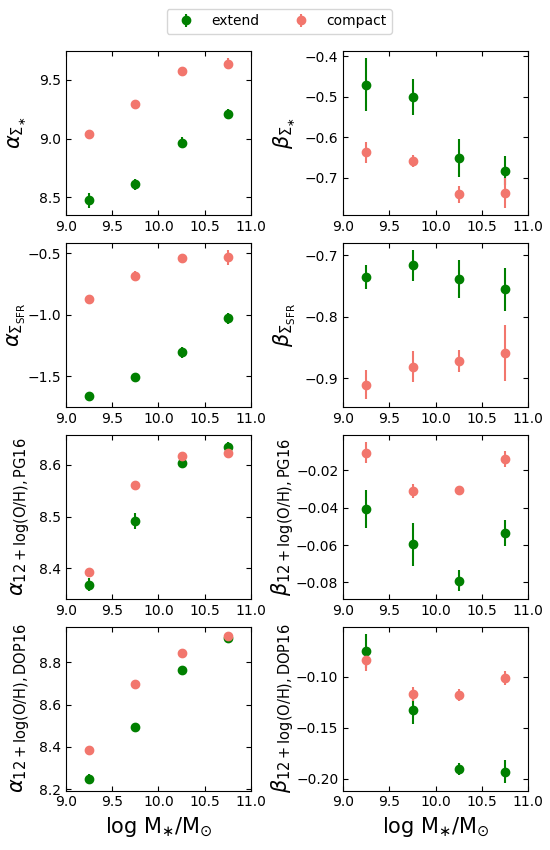}
	\end{center}
	\caption{The fitting coefficients for \sigstar, \sigsfr, 
 \loh$_{\rm PG16}$, and \loh$_{\rm DOP16}$ profiles as a function of stellar mass. Compact and extended galaxies are shown in coral and green dots accordingly with their errors.}
	\label{fig:slope}    
\end{figure}

\begin{deluxetable*}{lllllllll}
    \tablewidth{0pt}
    \tablecolumns{9}
    \tablecaption{The best-fitting coefficients in the linear relation $y=\alpha+\beta x$ for different radial profiles.} 
    \tablehead{\multicolumn{1}{c}{Mass-Size Bin} & \multicolumn{2}{c}{\sigstar} & \multicolumn{2}{c}{\sigsfr} & \multicolumn{2}{c}{12+log(O/H)$_{\rm PG16}$} & \multicolumn{2}{c}{12+log(O/H)$_{\rm DOP16}$}\\
    \colhead{} & \colhead{$\alpha$} & \colhead{$\beta$} & \colhead{$\alpha$} & \colhead{$\beta$} & \colhead{$\alpha$} & \colhead{$\beta$} & \colhead{$\alpha$} & \colhead{$\beta$}}
    \startdata
9.0-9.5(extended) & -1.66$\pm$0.02& -0.74$\pm$0.02&  8.47$\pm$0.06 & -0.47$\pm$0.06 & 8.37$\pm$0.01& -0.04$\pm$0.01 & 8.25$\pm$0.02& -0.08$\pm$0.02 \\
9.5-10.0(extended) &  -1.51$\pm$0.03& -0.72$\pm$0.02& 8.61$\pm$0.05 & -0.50$\pm$0.04 & 8.49$\pm$0.01& -0.06$\pm$0.01 & 8.49$\pm$0.02& -0.13$\pm$0.01 \\
10.0-10.5(extended) &  -1.30$\pm$0.04& -0.74$\pm$0.03& 8.96$\pm$0.05 & -0.65$\pm$0.05 & 8.60$\pm$0.01& -0.08$\pm$0.01 & 8.76$\pm$0.01& -0.19$\pm$0.01 \\
10.5-11.0(extended) &  -1.03$\pm$0.04&  -0.76$\pm$0.04& 9.21$\pm$0.04 & -0.68$\pm$0.03 &  8.64$\pm$0.01& -0.05$\pm$0.01 & 8.92$\pm$0.01& -0.19$\pm$0.01 \\
9.0-9.5(compact) &  -0.87$\pm$0.03& -0.91$\pm$0.02&  9.04$\pm$0.03 & -0.64$\pm$0.03 & 8.39$\pm$0.01&  -0.01$\pm$0.01 & 8.38$\pm$0.01& -0.08$\pm$0.01 \\
9.5-10.0(compact) & -0.68$\pm$0.03& -0.88$\pm$0.02&  9.29$\pm$0.02 & -0.66$\pm$0.01 & 8.56$\pm$0.01&  -0.03$\pm$0.00 & 8.70$\pm$0.01& -0.11$\pm$0.01 \\
10.0-10.5(compact) &  -0.54$\pm$0.02& -0.87$\pm$0.02& 9.58$\pm$0.03 & -0.74$\pm$0.02 & 8.62$\pm$0.00& -0.03$\pm$0.00 & 8.85$\pm$0.01& -0.12$\pm$0.01 \\
10.5-11.0(compact) &  -0.53$\pm$0.06& -0.86$\pm$0.05& 9.63$\pm$0.05 & -0.74$\pm$0.04 & 8.62$\pm$0.01& -0.01$\pm$0.00 & 8.93$\pm$0.01& -0.10$\pm$0.01
    \enddata
\end{deluxetable*}     \label{tbl:slope} 

\begin{deluxetable*}{lrrrrrr} 
    \tablewidth{0pt}
    \tablecolumns{9}
    \tablecaption{The best-fitting coefficients in the linear relation $y=\alpha+\beta x$ for resolved SFMS and MZRs.} 
    \tablehead{\multicolumn{1}{c}{Mass-Size Bin} & \multicolumn{2}{c}{\sigstar-sSFR} & \multicolumn{2}{c}{\sigstar-12+log(O/H)$_{\rm PG16}$} & \multicolumn{2}{c}{\sigstar-12+log(O/H)$_{\rm DOP16}$}\\
    \colhead{} & \colhead{$\alpha$} & \colhead{$\beta$} & \colhead{$\alpha$} & \colhead{$\beta$} & \colhead{$\alpha$} & \colhead{$\beta$}}
    \startdata
9.0-9.5(extended) &   -13.2$\pm$2.0 &     0.35$\pm$0.25&     6.90$\pm$0.08&     0.18$\pm$0.01&      5.78$\pm$0.10&     0.30$\pm$0.01 \\
9.5-10.0(extended) &   -12.4$\pm$1.8 &     0.26$\pm$0.23&     6.96$\pm$0.05&      0.18$\pm$0.01&     5.43$\pm$0.14&     0.36$\pm$0.02 \\
10.0-10.5(extended) &   -9.9$\pm$1.3 &     -0.05$\pm$0.16&     7.33$\pm$0.09&     0.14$\pm$0.01&      5.95$\pm$0.24&     0.32$\pm$0.03 \\
10.5-11.0(extended) &   -11.0$\pm$1.1 &     0.08$\pm$0.13&      7.92$\pm$0.15&     0.08$\pm$0.02&      6.34$\pm$0.29&     0.28$\pm$0.04 \\
9.0-9.5(compact) &   -14.3$\pm$0.7 &     0.49$\pm$0.08&     8.07$\pm$0.08&     0.04$\pm$0.01&      6.96$\pm$0.22&     0.16$\pm$0.03 \\
9.5-10.0(compact) &   -12.7$\pm$0.5 &     0.29$\pm$0.06&     8.05$\pm$0.04&     0.06$\pm$0.01&      7.14$\pm$0.12&     0.17$\pm$0.01 \\
10.0-10.5(compact) &   -11.8$\pm$0.5 &     0.17$\pm$0.05&     8.24$\pm$0.03&     0.04$\pm$0.01&      7.40$\pm$0.09&     0.15$\pm$0.01 \\
10.5-11.0(compact) &   -11.5$\pm$1.0 &     0.14$\pm$0.11&      8.45$\pm$0.05&     0.02$\pm$0.01&      7.63$\pm$0.13&     0.13$\pm$0.01 \\
\enddata
\end{deluxetable*}     \label{tbl:relation_slope}

\begin{figure*}
	\begin{center}
            \includegraphics[width=0.8\textwidth]{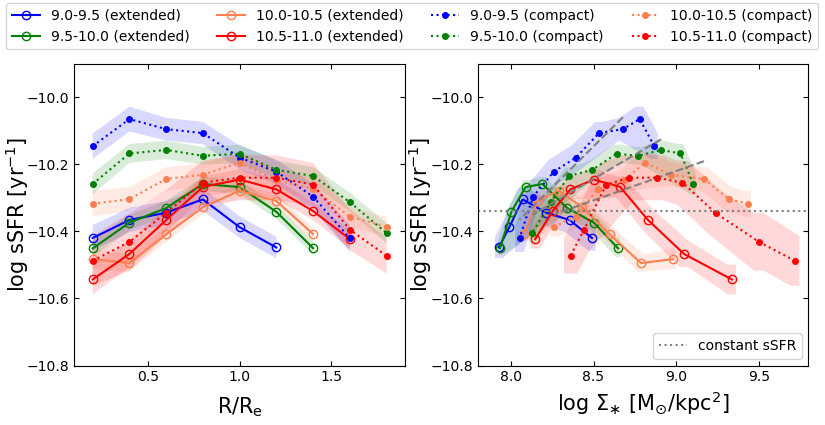}
	\end{center}
	\caption{Radial profiles of sSFR (left) and the stellar mass density (\sigstar)-sSFR relationship (right)  in different mass-size bins. The color of the lines represents different stellar mass bins. In each mass bin, dotted lines with filled dots are the median profile from compact galaxies, while solid lines with empty circles are the median profiles from extended galaxies. The shaded region represents the error on the median profile. In the right panel, grey dashed lines are the linear fits to the \sigstar-sSFR relationships (only the fits for three lower mass bins in compact galaxies are shown for clarity.)} A constant sSFR is indicated as the dotted grey line \citep{Hsieh-17}. 
	\label{fig:rSFMS}    
\end{figure*}

\begin{figure*}
	\begin{center}
            \includegraphics[width=0.8\textwidth]{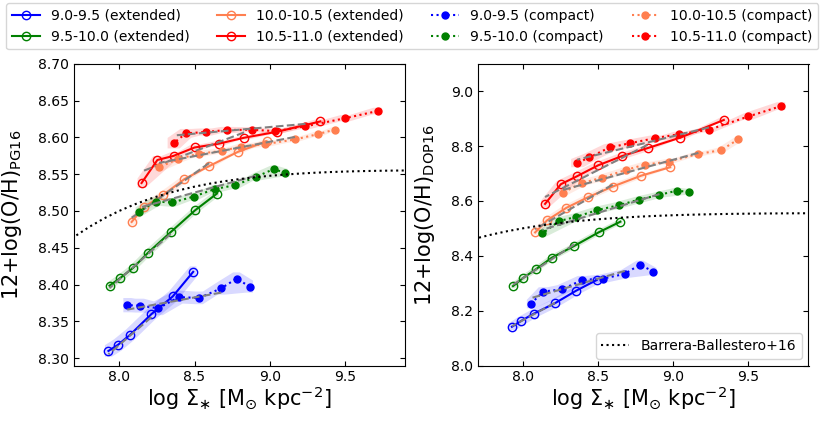}
	\end{center}
	\caption{The metallicity as a function of stellar mass density (\sigstar) in different mass-size bins. Both \loh$_{\rm PG16}$ (left) and \loh$_{\rm DOP16}$ (right) are shown. The color of the lines represents different stellar mass bins. Symbols are the same as Figure \ref{fig:rSFMS}. Grey dashed lines are the linear fits to the \sigstar-12+log(O/H) relationships. The best fit resolved mass-metallicity relation in \citet{Barrera-Ballesteros-16} is indicated by the dotted line. }
	\label{fig:rMZR}    
\end{figure*}

\begin{figure*}
	\begin{center}
		\includegraphics[width=0.8\textwidth]{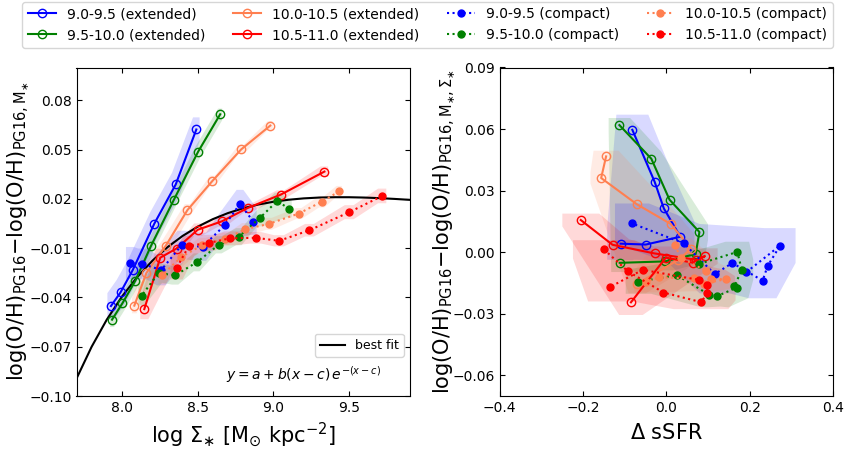}
            \includegraphics[width=0.8\textwidth]{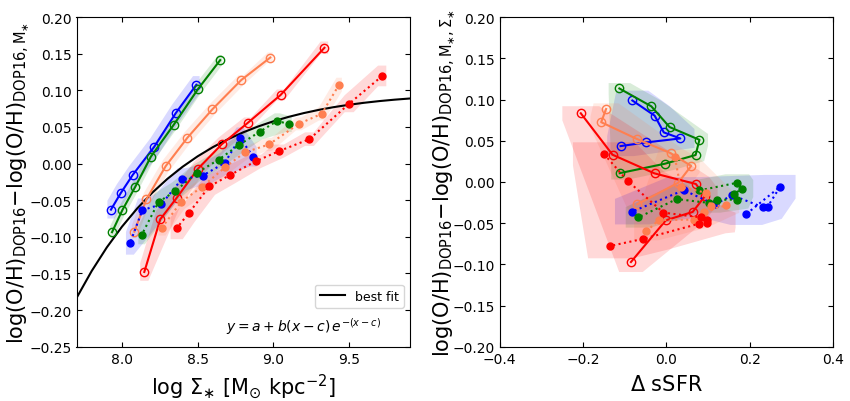}
	\end{center}
	\caption{Upper-left: The metallicity residuals at fixed \mstar\ as a function of local stellar mass density \sigstar. The best-fit curve is shown as a solid line, following the formula provided in the lower-right corner. Upper-right: The residuals of metallicity at fixed \mstar\ and \sigstar\ as a function of \dssfr. Symbols are the same as Figure \ref{fig:rSFMS}. See the text for more details on how these metallicity residuals are calculated. Bottom: Same as the top panels, but for \loh$_{\rm DOP16}$.}
	\label{fig:dZ_dSFR}    
\end{figure*}

\section{Comparison between Compact and Extended Galaxies} \label{sec:comparison}

\subsection{Global Properties}
\label{sec:comparison:global}


We start our analysis by exploring the global properties of our selected galaxies across the mass-size plane. Figure \ref{fig:global} and \ref{fig:global2} illustrates \dlsfr, log M$_{\rm HI}$, \Av, \loh$_{\rm tot}$, $\lambda_{R_{\rm e}}$, $\sigma_{\ast, cen}$, and Sersic $n$ as a function of \dlre. The radius residuals at fixed stellar mass, \dlre, are calculated from the residuals of the mass-size relation for star-forming galaxies in \citet{Shen-03}. \dlsfr\ is calculated as the residuals of SFR at fixed stellar mass according to the SFMS in \citet{Renzini-Peng-15}. The global atomic hydrogen mass (M$_{\rm HI}$) data are taken from the \HI\ follow-up observations being carried out by \citet{Masters-19}. The dust extinction \Av\ is derived from the H${\alpha}$/H${\beta}$ ratio, where H${\alpha}$ and H${\beta}$ represent the total integrated flux of the Gaussian fit to each emission line within the entire MaNGA field-of-view. The \loh$_{\rm tot}$ is the gas-phase metallicity calculated using the PG16 metallicity calibrator (or DOP16 calibrator) based on the total integrated emission lines within the entire galaxy. The stellar spin parameter $\lambda_{R_{\rm e}}$ is taken from \citet{Zhu-23}, which provides a catalog of dynamical properties derived from the JAM method. Stellar velocity dispersion data are taken from the DAP maps. The Sersic index $n$ is measured from the MaNGA PyMorph catalog \citet{Dominguez_Sanchez-22}, which performed one Sersic fit to the 2D surface brightness profiles of the final MaNGA DR17 galaxy sample. In each panel, we measure the linear correlation between the two sets of data, with the Pearson correlation coefficient ($r$) and the $p-$value listed in the top-right corner.

The first row of Figure \ref{fig:global} shows that the SFRs between compact and extended galaxies at fixed stellar mass are similar, consistent with the lack of a significant correlation between SFR and size in \citet{Lin-20}. The median \dlsfr\ for compact and extended galaxies are represented by larger symbols in each panel, and they fall within 1-2$\sigma$ of each other. This lack of correlation is further supported by the Pearson correlation coefficients ($r$), which are close to zero in these panels. While \citet{Stephenson-24} report a non-monotonic relation between SFR and size at fixed stellar mass, our sample shows almost no correlation.


The similar SFRs between compact and extended galaxies suggest they have comparable amounts of molecular gas, given the strong correlation between SFR and molecular gas. This aligns with the observed weak correlation between molecular depletion time and radius \citep{Saintonge-17, Tacconi-18}. However, the second row in Figure \ref{fig:global} shows that compact galaxies have lower \HI\ content. While the correlation coefficients are moderate ($r\sim0.3$), the low $p-$values ($<0.001$) indicate a significant correlation.  This result is actually expected, as global \HI\ mass is known to correlate strongly with disk size \citep{Wang-16, Yesuf-19}.

Despite having a lower total gas fraction (comparable H$_2$ plus less \HI\ gas), compact galaxies reveal higher dust extinction due to their elevated gas (and dust) surface density. Considering the variation of \dlre, approximately 0.3 dex from compact ($-$0.15 dex) to extended (+0.15 dex) galaxies, we anticipate an increase of 0.6 dex in gas (and dust) surface density, as well as an $\Delta$\Av$=$0.6 dex if we assume a foreground screen of dust. However, the observed changes in \Av\ are much less than that, only $\sim$0.25 dex higher than extended galaxies. We will discuss this later in Section \ref{sec:discuss:SFE}. 

Next, we show that compact galaxies have higher metallicity compared with extended galaxies, as demonstrated in the fourth and fifth rows. This finding does not change by making use of the DOP16 calibrator, although it does lead to shifts in the absolute values. The correlation is significant (with $p-$values $<$ 0.001) and consistent with the previous results derived from the central 3\arcsec\ SDSS spectra \citep{Ellison-08a}.

Last, we show that compact galaxies have lower angular momentum $\lambda_{R_{\rm e}}$ and higher central stellar velocity dispersion $\sigma_{\ast, cen}$ compared with extended galaxies. We also compare their light profiles by plotting the Sersic index $n$ in the last row. Compact galaxies reveal higher concentration than extended ones (higher Sersic $n$ in single-component fits), with the difference becoming more pronounced at the high mass end. All of the $p-$values indicate significant correlations.

\subsection{Radial Profiles}
\label{sec:comparison:profiles}

We display the median radial profiles of \sigsfr, \sigstar, \loh$_{\rm PG16}$ and \loh$_{\rm DOP16}$ in Figure \ref{fig:radprofiles}. The color of the lines represents different stellar mass bins. In each mass bin, dotted lines with filled dots are the median profile from compact galaxies, while solid lines with empty circles are the median profiles from extended galaxies. There are clear offsets in the y-axis of these profiles, with more massive galaxies tending to have higher \sigstar, higher \sigsfr, and higher metallicity.

One concern regarding the comparison of profiles between compact and extended galaxies is the beam-smearing effect caused by the Point Spread Function (PSF). This effect leads to flattening of the profiles in the very central regions. As we show the profile in units of $R$/\re, compact galaxies with smaller \re\ suffer relatively severe beam-smearing effects, since their \re/PSF ratios are small: less resolution elements per \re\ \citep{Belfiore-17}. See Appendix \ref{app:psf} for a more detailed evaluation for beam-smearing effect.

Here we quantified the slope of the profiles by excluding the central regions, following the suggestion in \citet{Belfiore-17} to minimize the beam-smearing effect on the slope measurements. We fit the profile in $R$ $>$ 0.5 \re\ using a straight-line model $y=\alpha+\beta x$, where $x$ is the radius in units of $R$/\re. The grey dashed lines in Figure \ref{fig:radprofiles} are best fitting lines to the median profiles in each mass-size bin. The fitting coefficients are presented in Table \ref{tbl:slope}. Figure \ref{fig:slope} shows the coefficients of the fitting for \sigstar, \sigsfr, \loh$_{\rm PG16}$ and \loh$_{\rm DOP16}$ profiles as a function of stellar mass, respectively. 

Both compact and extended galaxies have steeper \sigstar\ gradients towards the high mass end, suggesting a higher concentration in massive galaxies. Extended galaxies have similar \sigsfr\ gradients across different mass bins, while less massive compact galaxies exhibit steeper \sigsfr\ gradients. Compared with extended galaxies, although compact galaxies could suffer severer beam-smearing effect, compact ones actually have steeper slopes (coefficient $\beta$) in both \sigsfr\ and \sigstar, with the difference in their \sigsfr\ slopes being more prominent. Therefore, we suspect net negative gradients in sSFR ($=$\sigsfr/\sigstar) and its variation with stellar mass. 

Additionally, it is also shown that compact and extended galaxies have different metallicity slopes. Extended galaxies tend to have steeper gradients at fixed stellar mass, whereas compact galaxies display flatter slopes. This trend is seen in both \loh$_{\rm PG16}$ and \loh$_{\rm DOP16}$ profiles. The result is consistent with \citet{Boardman-21}, who also found that metallicity gradients vary with \re\ using the MaNGA datasets. The overall metallicity gradients reveal a mass dependence, with gradients steepening with increasing stellar mass and become flatten in central regions in the highest mass bin \citep{Belfiore-17}. 

In summary, our results reveal distinct profiles in \sigstar, \sigsfr, and \loh\ for compact and extended galaxies. For \sigstar\ and \sigsfr, there is not only higher surface densities in compact galaxies, but the gradients also change with galaxy size. In the next section, we will investigate whether compact and extended galaxies follow similar relations at fixed \sigstar.

\section{Scaling Relations}
\label{sec:scaling_relations}

\subsection{Resolved Star Formation Main Sequence}

In Figure \ref{fig:rSFMS}, we present the median radial profiles of 
specific star formation rate (sSFR), calculated as the ratio of star formation rate surface density (\sigsfr) to stellar surface density (\sigstar) in each annulus. The right panel shows the relationship between sSFR and the stellar mass density (\sigstar) for different mass-size bins. We fit the \sigstar-sSFR relation using a linear model $y=\alpha+\beta x$ to characterize the differing trends between compact and extended galaxies. Only data points with $r > 0.5$\re\ are included in the fit, and the coefficients are listed in Table \ref{tbl:relation_slope}. The lines are color-coded to present different stellar mass bins, with dotted lines and filled circles indicating compact galaxies and solid lines with empty circles representing extended galaxies.

For extended galaxies, the sSFR profiles show slightly lower values in the inner region or are closed to a relatively constant, as shown by the dashed grey line in the right panel of Figure \ref{fig:rSFMS}, consistent with previous studies \citep{Cano-Diaz-16, Hsieh-17}. 

In contrast, compact galaxies have higher sSFR in their inner regions, characterized by negative sSFR gradients. These steep gradients are more pronounced in the lower mass bins. As shown in the right panel of Figure \ref{fig:rSFMS}, while the slope in the \sigstar-sSFR relation approaches zero at the massive end, it becomes significantly positive in the lowest mass bin ($\beta=0.49\pm0.08$). This supports our previous expectation, as the \sigsfr\ profiles in compact galaxies display steeper gradients than the \sigstar\ profiles (See Section \ref{sec:comparison:profiles}). At fixed \sigstar, compact galaxies have higher sSFR compared to regions with the same \sigstar\ in more massive extended galaxies.

\subsection{Resolved Mass-Metallicity Relation}

In Figure \ref{fig:rMZR}, we plot the resolved mass-metallicity relation in different mass-size bins using \loh$_{\rm PG16}$ and \loh$_{\rm DOP16}$ separately. Symbols are the same as Figure \ref{fig:rSFMS}. Grey dashed lines represent the linear fit for the \sigstar-12+log(O/H) relation in each mass-size bin (See Table \ref{tbl:relation_slope}).  For comparison, the resolved mass-metallicity relation from \citet{Barrera-Ballesteros-16}, which uses the {\tt O3N2} metallicity calibrator, is shown as the dotted line. 

At a given stellar mass bin, compact and extended galaxies do not follow the same trend in the resolved MZ relation. For extended galaxies, the slope of the MZ relation (\sigstar-\loh$_{\rm PG16}$) ranges from 0.08 to 0.18, with a median value of 0.14$\pm$0.01. In contrast, the slope for compact galaxies is shallower, ranging from 0.02 to 0.06, with a median value of 0.04$\pm$0.01. The difference is also seen in the \sigstar-\loh$_{\rm DOP16}$ plot. 

Thus, extended galaxies have steeper trends in the mass-metallicity relation, with their outer regions showing lower \sigstar\ and \loh. While the metallicity in compact galaxies do not vary significantly as a function of \sigstar, consistent with the flat metallicity profiles shown in Section \ref{sec:comparison:profiles}.

\subsection{Metallicity-$\Sigma_{\rm SFR}$ Relation}

In the previous sections, we have demonstrated that compact galaxies have higher sSFR and different metallicity gradients compared with extended galaxies. Given the anti-correlation between SFR surface density and metallicity found in several studies, this motivate us to consider whether a relatively lower metallicity is related to the higher sSFR in compact galaxies, which could imply a more efficient metal-poor gas inflows or metal-rich gas outflows in compact galaxies. 

One issue in comparing the metallicity and SFR is that both of them correlate with global stellar mass and local stellar mass density. Therefore, it is necessary to remove the mass dependence and then compare the residuals of the two quantities. We describe our approach for removing the mass dependence in the following.

In the upper-left panel of Figure \ref{fig:dZ_dSFR}, we plot the resolved mass-metallicity relation again, but this time the metallicity profiles are shifted by subtracting the global values corresponding to the given mass-size bins, as shown in the fourth and fifth rows of Figure \ref{fig:global} for \loh$_{\rm PG16}$ and \loh$_{\rm DOP16}$, respectively. We re-fit the resolved mass-metallicity relation by removing the global \mstar\ dependence. For the fitting, we adopt the form used in \citet{Barrera-Ballesteros-16}: $y=a+b(x-c)e^{-(x-c)}$, where $x=$\,log\,\sigstar\ in units of \msun\,pc$^{-2}$ and $y$ is the metallicity residual after removing the global mass dependence. The best-fitted coefficients depend on which metallicity calibration we adopted. For \loh$_{\rm PG16}$, the fitted coefficients are $a=0.00\pm0.03$, $b=0.06\pm0.06$, and $c=2.44\pm0.55$. For \loh$_{\rm DOP16}$, the best-fitted coefficients are $a=0.09\pm0.07$, $b=0.01\pm0.06$, and $c=4.0\pm1.8$. 

We then calculate the metallicity residuals at fixed \mstar\ and \sigstar\ according to our best-fitted lines. We plot the residuals of metallicity at fixed \mstar\ and \sigstar\ versus \dssfr\ in the upper-right panel of Figure \ref{fig:dZ_dSFR}. The \dssfr\ is also a quantity that removes the global mass and local mass density dependences in SFR. We find there is no obvious correlation between the metallicity residuals and \dssfr. This result remains valid if we use different metallicity calibrator. The relations using DOP16 are shown in the bottom panels of Figure \ref{fig:dZ_dSFR}. 

The weak correlation between metallicity residuals and \dssfr\ at fixed \mstar\ and \sigstar\ does not change if we use different ways to remove the global mass dependence. For example, one could use metallicity within \re\ or central metallicity to characterize the global mass-metallicity relation. This would only shift the zero-points of each mass-size bin on the $y$-axis, but does not affect the overall result.

\begin{figure*}
	\begin{center}
		\includegraphics[width=\textwidth]{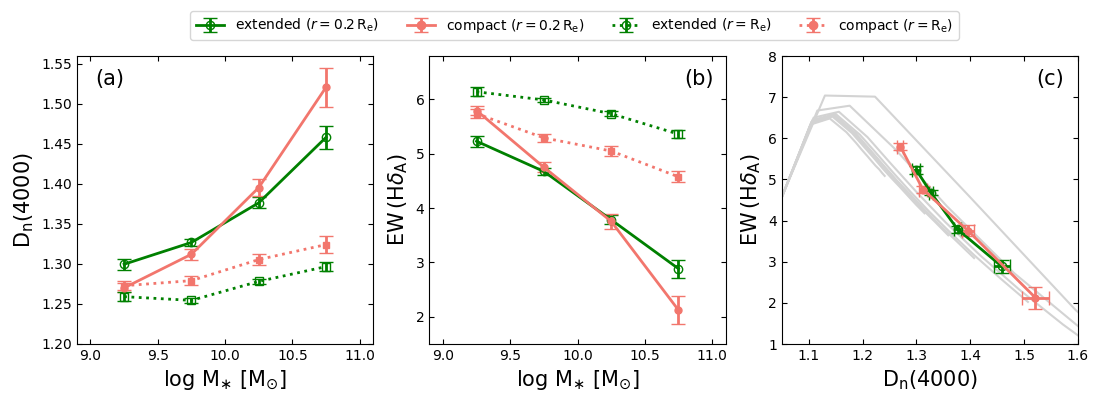}
	\end{center}
	\caption{(a) The changes of D$_{n}$(4000) as a function of stellar mass with corresponding errors. The central values change with different stellar masses are shown in solid lines, the trend at \re\ are shown in dashed lines. Compact and extended galaxies separated into filled coral dots and empty green circles accordingly. (b) The changes of EW(H$\delta$) as a function of stellar mass. (c) D$_{n}$(4000) versus EW(H$\delta$) diagram. The grey solid lines are the predictions of Bruzual \& Charlot (2003) model with declining star formation history. }
	\label{fig:sfh}    
\end{figure*}

\section{Discussion} \label{sec:discuss}

\subsection{What cause the higher sSFR in inner regions of compact galaxies?} \label{sec:discuss:SFE}

In this work, we observed distinct sSFR slopes between compact and extended galaxies. Extended galaxies exhibit a flat sSFR profile, consistent with previous result, whereas compact galaxies display a negative sSFR gradient, with higher sSFR values in their inner regions compared to the outskirts. 

Recent studies indicated that the scatter in resolved SFMS is attributed to the variations in star formation efficiency (SFE) or molecular gas fraction \citep{Ellison-21, Pessa-21}. Since we have shown that compact and extended galaxies have similar global SFR, we expect that their molecular gas fractions are similar. We further discuss their resolved star formation efficiency (SFE) profiles in the following.

From observation, the ALMA-MaNGA QUEnch and STar formation (ALMaQUEST) survey \citep{Lin-19, Ellison-24} has provided CO observations for 66 MaNGA galaxies. Although many of these galaxies are starburst or green valley galaxies, there are still 12 star-forming galaxies with stellar mass larger than 10$^{10}$\msun\ that can be matched with our sample (including 4 compact galaxies and 8 extended galaxies). Both star formation rate surface density (\sigsfr) and molecular gas surface density (\sightwo) maps are available in \citet{Ellison-24}, where a constant CO-to-H$_{\rm 2}$ conversion factor ($\alpha_{\rm CO}$) is adopted. We calculate the median molecular gas SFE (\sigsfr/\sightwo) profiles for compact and extended galaxies separately and find similar values ($\sim 10^{-9}\,$ yr$^{-1}$) of SFE in the central regions of compact and extended galaxies.

An alternative method to estimate gas content is based on dust absorption. Despite relatively large uncertainties associated with this approach \citep{Boquien-13}, it can be simply applied to a large sample using optical data \citep{Brinchmann-13}. We use the recent calibration in \citet{Barrera-Ballesteros-20} to estimate gas profiles from \Av\ profiles. They established the \siggas-\Av\ relation as: \siggas $= (27 \pm 5)$ \Av, where \siggas $=$ \sightwo\ $+$ \sighone\ (including the mass contribution from Helium) and \Av\ represents the gas-phase dust extinction derived from the Balmer decrement. We then calculated the total gas SFE profiles by dividing \sigsfr\ by \siggas. Compared with extended galaxies at fixed stellar mass, compact galaxies have significant higher \sigsfr, but the \Av\ does not increase proportionally (see the median profiles in Appendix \ref{app:profiles}). Thus, central regions in compact galaxies would have higher SFE ($\sim 10^{-8}\,$ yr$^{-1}$) in this approach, which seems to contradict the findings from ALMaQUEST observations mentioned in the previous paragraph.

There are two ways to reconcile this discrepancy. First, the accuracy of gas mass measurements heavily relies on the CO-to-H$_{\rm 2}$ conversion factor ($\alpha_{\rm CO}$). Recent studies have revealed strong $\alpha_{\rm CO}$ variations on metallicity, SFR surface density, stellar surface density, gas intensity, and velocity dispersion \citep{Accurso-17, He-24, Chiang-24}. For instance, adopting a velocity dispersion dependent $\alpha_{\rm CO}$ as demonstrated in \citet{Teng-24} would result in a significant increase in SFE in high surface density regimes (such as the central regions of compact galaxies). In contrast, using a constant or metallicity-based $\alpha_{\rm CO}$ would produce a more uniform SFE distribution. Second, the relation between \siggas-\Av\ is motivated by the optical depth (\Av) and gas column density ($N_{\rm H}$) relationship observed in the Milky Way, which assumes a foreground screen dust geometry and solar metallicity. However, nearby galaxies have a more complex dust geometry and a broader range of metallicities, potentially deviating from this linear relationship. Detailed turbulent interstellar medium (ISM) modeling shows that while \Av\ generally increases with dust column density, it begins to saturate in highly clumpy ISM \citep{Seon-Draine-16}, indicating that a simple linear relation between \siggas\ and \Av\ may lead to an underestimation of the actual amount of dust (or gas) \citep{Qin-19, Lu-22}.

In summary, there are indeed indications of higher SFE in compact galaxies compared with extended galaxies. However, an accurate assessment of SFE profiles in various environments requires further investigation, which is beyond the scope of this paper. We postpone this analysis to future work.

\subsection{The lack of correlation between SFR and metallicity}

The correlation between SFR and metallicity has been studied extensively but remains controversial. The physical connection between metallicity and SFR is expected in the gas inflow scenario that the accretion of metal-poor gas from the intergalactic medium or through mergers can boost star formation and dilute the metallicity of a galaxy, causing the anti-correlation between SFR and Z \citep{Mannucci-10, Lara-Lopez-10}. But other studies also argue that the lower metallicity can be quickly smoothed over as star formation continues \citep{Torrey-12, De_Los_Reyes-15}. Despite some evidence for the dependence of metallicity on SFR or sSFR \citep{Andrews-Martini-13, Curti-17, Cresci-19, Curti-20}, there are several concerns that the observed FMR is driven by the selection of metallicity calibration \citep{Telford-16, Teklu-20}, the aperture effects \citep{Sanchez-17}, improper subtraction of the stellar mass dependence on metallicity \citep{Barrera-Ballesteros-17}, or the contribution of extremely starbursts \citep{Ellison-18}. 

In this work we assess the correlation between SFR residuals and metallicity residuals at fixed global mass and local mass surface density. Not obvious correlation is seen. A similar approach has been used in previous works such as \citet{Barrera-Ballesteros-16, Gao-18, Boardman-22}. These studies also confirmed that the combination of local stellar mass surface density and total stellar mass is sufficient to capture the overall mass–size trend, with no obvious dependence of SFR at fixed \mstar\ and \sigstar.

A different way to explore the connection between SFR and metallicity involves removing their radial profiles and then comparing the residuals. \citet{Sanchez-Menguiano-19} studied the relation between SFR and metallicity after subtracting their radial profiles. They observed that most galaxies exhibit an anti-correlation between SFR and metallicity, with the strongest anti-correlation appeared in the more metal-poor galaxies. Similarly, \citet{Wang-Lilly-21} removed the global mass and the radial dependencies and found a negative correlation between SFR and gas metallicity. Given our findings that the radial slopes of both SFR and metallicity depend on both global mass and galaxy size, it is plausible that the residuals calculated based on radial profiles could inherit a certain dependency. This could potentially lead to their observed correlation between SFR and metallicity.

\subsection{Implication for different SFHs in compact galaxies}

Now we turn to discuss the star formation history in compact and extended galaxies. 

Figure \ref{fig:sfh}a shows the changes in \dindex\ as a function of stellar mass. As stellar mass increase, central \dindex\ values become larger, indicating older stellar populations in more massive galaxies. For comparison, we also plot the D$_{n}$(4000) at \re\ as a function of stellar mass as dotted lines. We observe that as the total mass increases, the stellar age in the outer disk is also getting old, albeit more gradually.

The interesting point arises when we separate the compact and extended galaxies. In central regions, compact galaxies are younger than extended galaxies at the low-mass end, but they are older than extended galaxies at the high-mass end. Conversely, If we compare their ages in the disk region, both compact and extended galaxies exhibit a parallel aging trend. This pattern is similarly shown in the equivalent width H$\delta$ (\ewhd) shown in Figure \ref{fig:sfh}b, except that a lower \ewhd\ indicates an older age.

Recent studies suggest distinct evolutionary pathways for compact and extended galaxies. \citet{Woo-Ellison-19} explored galaxies with different $\Sigma_{\rm 1kpc}$ and found that compact galaxies have higher sSFR in the center, which is consistent with our results since $\Sigma_{\rm 1kpc}$ and size are well correlated at fixed stellar mass. They proposed that compact galaxies experience more compaction events, while extended galaxies follow a more secular inside-out growth picture. Similar conclusions were drawn from the TNG50 simulation by \citet{Du-21} and \citet{Ma-24}, who found that compact and extended galaxies represent two distinct evolutionary trajectories. Compact galaxies tend to be bulge-dominated systems that develop their massive bulge components in the early stage and then grow their disk structures in a later phase.

We further plot the \dindex\ versus \ewhd\ diagram to probe the recent star formation history in compact and extended galaxies (Figure \ref{fig:sfh}c). The grey solid lines represent the model predictions of Bruzual \& Charlot (2003) with a declining star formation history. For clarity, we only plot the central values in this diagram. Our analysis reveals that, as stellar mass increases, compact galaxies initially exhibit a younger age and eventually transition to older populations compared to extended galaxies, but the trend in compact galaxies overlaps with that in extended galaxies. This suggests that they might follow a similar recent evolutionary track, but central regions in compact galaxies exhibit a more rapid increase in their old stellar populations. 

Combined with the higher sSFR and possible higher SFE in previous sections, our results indicate that the gas in the centers of compact galaxies is being rapidly consumed and then the average stellar population age are getting old quickly, suggesting a faster central growth in compact galaxies.

\section{Summary} \label{sec:summary}

In this study, we analyzed the radial profiles of \sigstar, \sigsfr, and \loh\ for a sample of $\sim$1240 nearby star-forming galaxies using integral field spectroscopy data from the MaNGA survey. We compared the global and resolved properties between compact and extended galaxies at fixed stellar mass. 

Our main findings can be summarized as follows.

(1) Compared with the global properties of extended galaxies at fixed \mstar, we find that compact galaxies show systematic differences in several parameters. Besides the expected lower angular momentum, compact galaxies have lower \HI\ gas fraction, higher dust extinction, higher metallicity, and higher concentration.

(2) We measure radial profiles of \sigstar\ and \sigsfr\ in compact and extended galaxies in units of $R$/\re. In addition to having higher surface densities, compact galaxies also show steeper gradients in \sigstar\ and \sigsfr, with the differences in \sigsfr\ profiles being more prominent. This results in negative gradients of sSFR in compact galaxies, indicating higher sSFR values in the inner regions compared to the outskirts. At fixed \sigstar, the central regions in compact galaxies show higher sSFR than the outer regions in more massive extended galaxies.

(3) We confirm that compact and extended galaxies have different metallicity profiles. At fixed \mstar, extended galaxies have steeper gradients, while compact galaxies show flatter slopes. 

(4) We investigate the connection between SFR and metallicity by calculating the SFR residuals and metallicity residuals at fixed global mass \mstar\ and local mass surface density \sigstar. No obvious correlation between SFR and gas metallicity has been found.

(5) The higher sSFR in the inner regions of compact galaxies implies that the inner gas in these galaxies would be quickly consumed. Using star formation history indicators like \dindex\ and \ewhd, we find that central old stellar populations are increase rapidly in compact galaxies, suggesting their central regions grow faster than the central parts of extended galaxies.


\begin{acknowledgments}

We thank the anonymous referee for comments that helped to improve the paper. This work was supported by the National Science Foundation of China (grant Nos. U1831205), the Natural Science Foundation of Shanghai (Grant No. 21ZR1474200), the Youth Innovation Promotion Association CAS (id. 2022260). LL thanks Meng Yang for discussing the transport of angular momentum in galaxies.

This work makes use of data from SDSS-IV. Funding for SDSS-IV has been provided by the Alfred P. Sloan Foundation and Participating Institutions. Additional funding towards SDSS-IV has been provided by the US Department of Energy Office of Science. SDSS-IV acknowledges support and resources from the Centre for High-Performance Computing at the University of Utah. The SDSS web site is www.sdss.org.

SDSS-IV is managed by the Astrophysical Research Consortium for the Participating Institutions of the SDSS Collaboration including the Brazilian Participation Group, the Carnegie Institution for Science, Carnegie Mellon University, the Chilean Participation Group, the French Participation Group, Harvard–Smithsonian Center for Astrophysics, Instituto de Astrofsica de Canarias, The Johns Hopkins University, Kavli Institute for the Physics and Mathematics of the Universe (IPMU)/University of Tokyo, Lawrence Berkeley National Laboratory, Leibniz Institut fur Astrophysik Potsdam (AIP), Max-Planck-Institut fur Astronomie (MPIA Heidelberg), Max-Planck-Institut fur Astrophysik (MPA Garching), Max-Planck-Institut fur Extraterrestrische Physik (MPE), National Astronomical Observatory of China, New Mexico State University, New York University, University of Notre Dame, Observatario Nacional/MCTI, The Ohio State University, Pennsylvania State University, Shanghai Astronomical Observatory, United Kingdom Participation Group, Universidad Nacional Autonoma de Mexico, University of Arizona, University of Colorado Boulder, University of Oxford, University of Portsmouth, University of Utah, University of Virginia, University of Washington, University of Wisconsin, Vanderbilt University and Yale University.

\end{acknowledgments}

%






\appendix

\section{Individual and Median profiles in different mass-size bins} 
\label{app:profiles}

Here we present the individual and the median profiles in different mass-size bins, including \sigsfr, \sigstar, \loh$_{\rm PG16}$, \loh$_{\rm DOP16}$, \Av, \dindex, \ewhd, and \ewha. The individual radial profiles are calculated from the 2-dimensional maps with PA and inclination angle from NSA photometric catalog.


\begin{figure*}
    \begin{center}
	\includegraphics[width=0.9\textwidth]{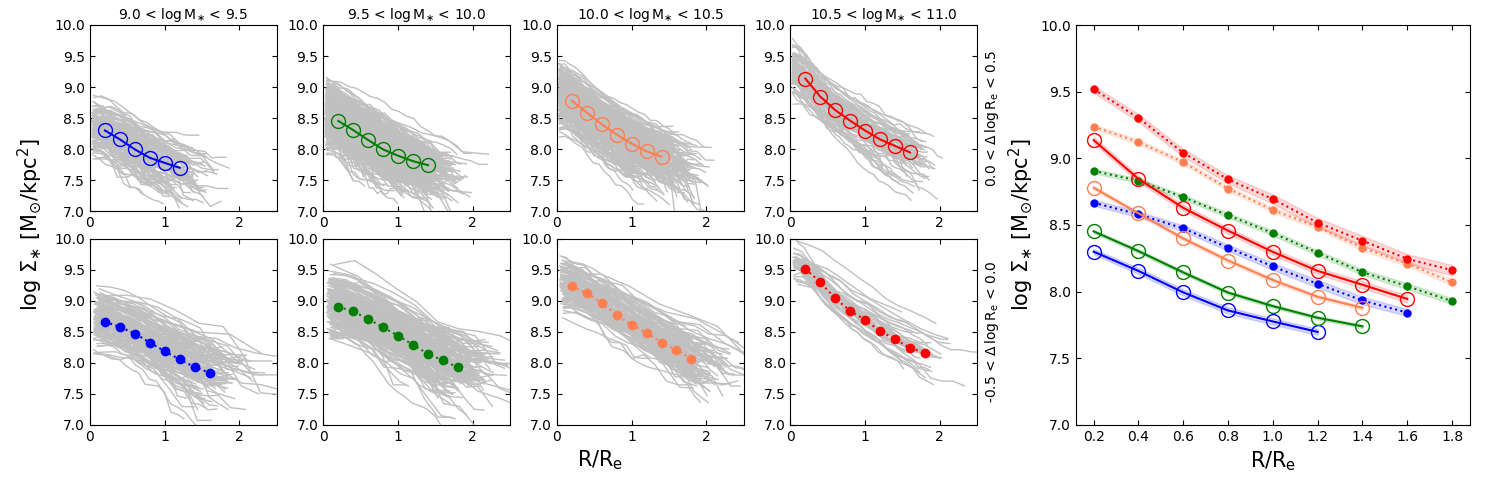}
        \includegraphics[width=0.9\textwidth]{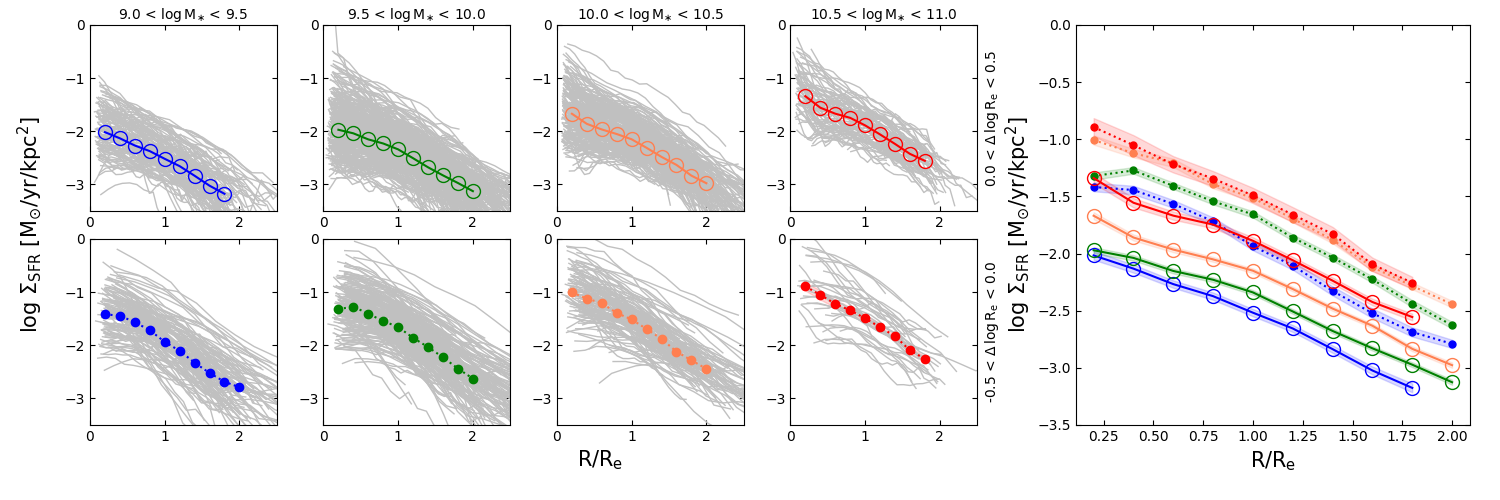}
        \includegraphics[width=0.9\textwidth]{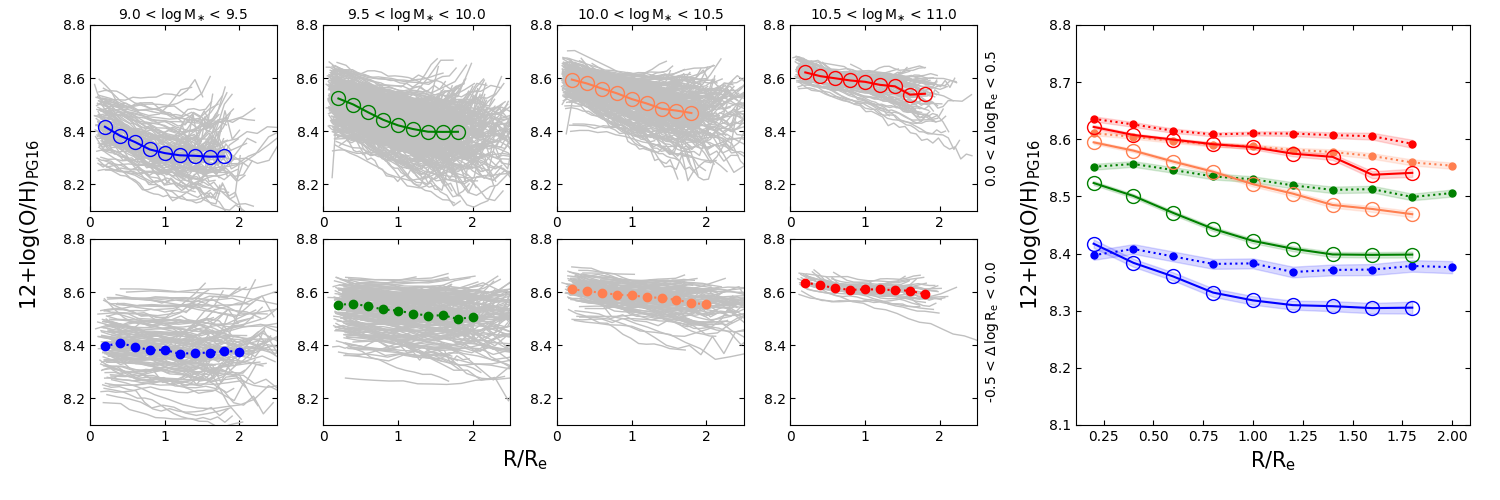}
        \includegraphics[width=0.9\textwidth]{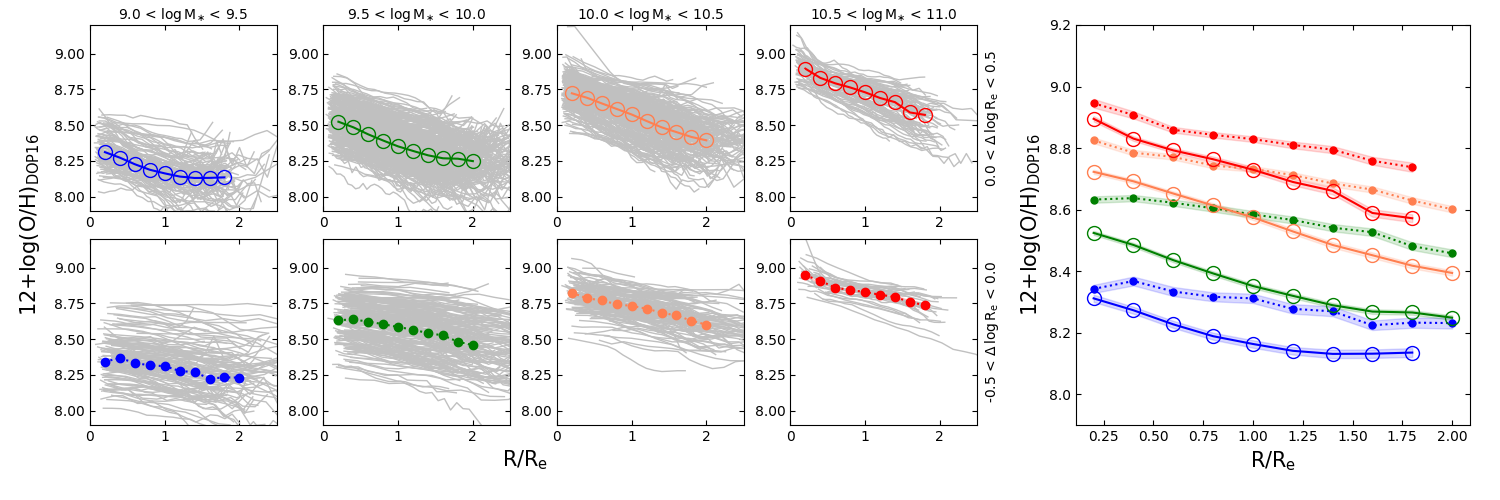}
    \end{center}
    \caption{Radial profiles of \sigstar, \sigsfr, \loh$_{\rm PG16}$, \loh$_{\rm DOP16}$, \Av, \dindex, \ewhd, and \ewha\ for each mass-size bin. Grey lines show the profiles for individual galaxies, with radius measured in steps of 0.2 $R_{\rm e}$ along the semi-major axis. The median profiles are color-coded by stellar mass, with solid lines representing extended galaxies and dotted line representing compact galaxies. The shaded regions indicate the errors on the median profiles.}
    \label{fig:sigstar_profiles}    
\end{figure*}

\begin{figure*}
    \centering
    \includegraphics[width=0.9\textwidth]{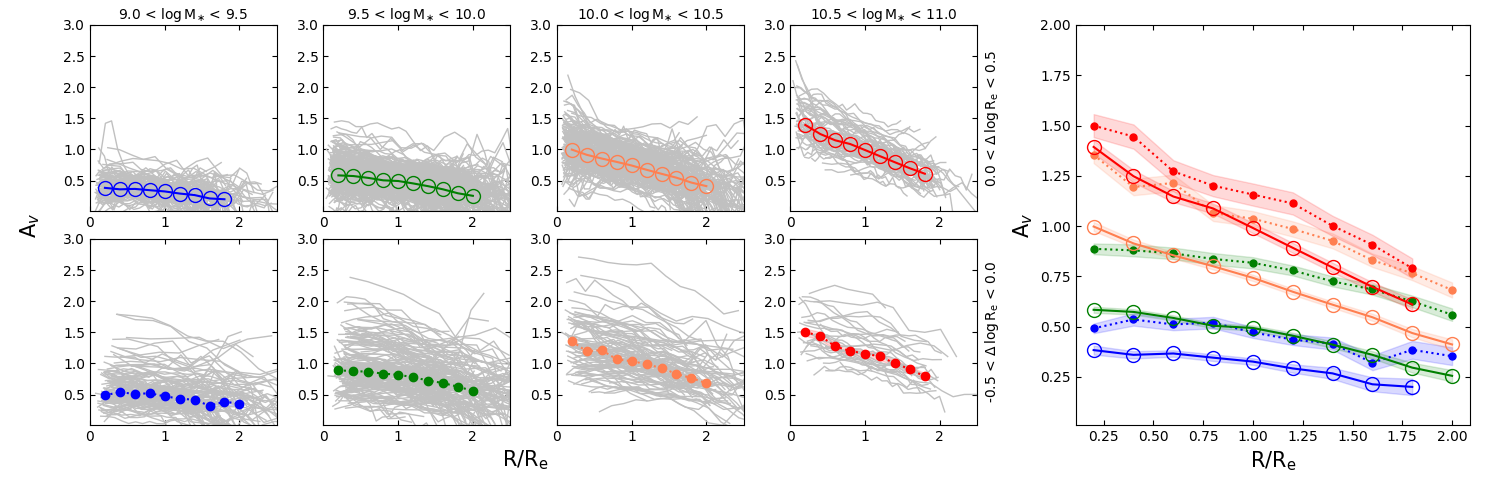}
    \includegraphics[width=0.9\textwidth]{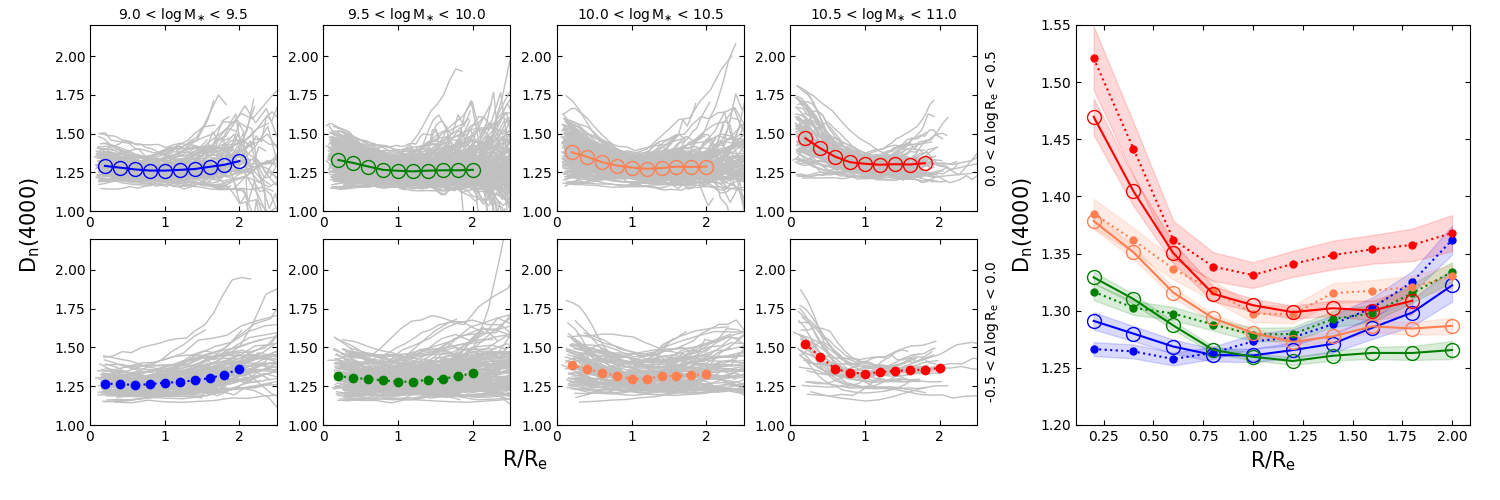}
    \includegraphics[width=0.9\textwidth]{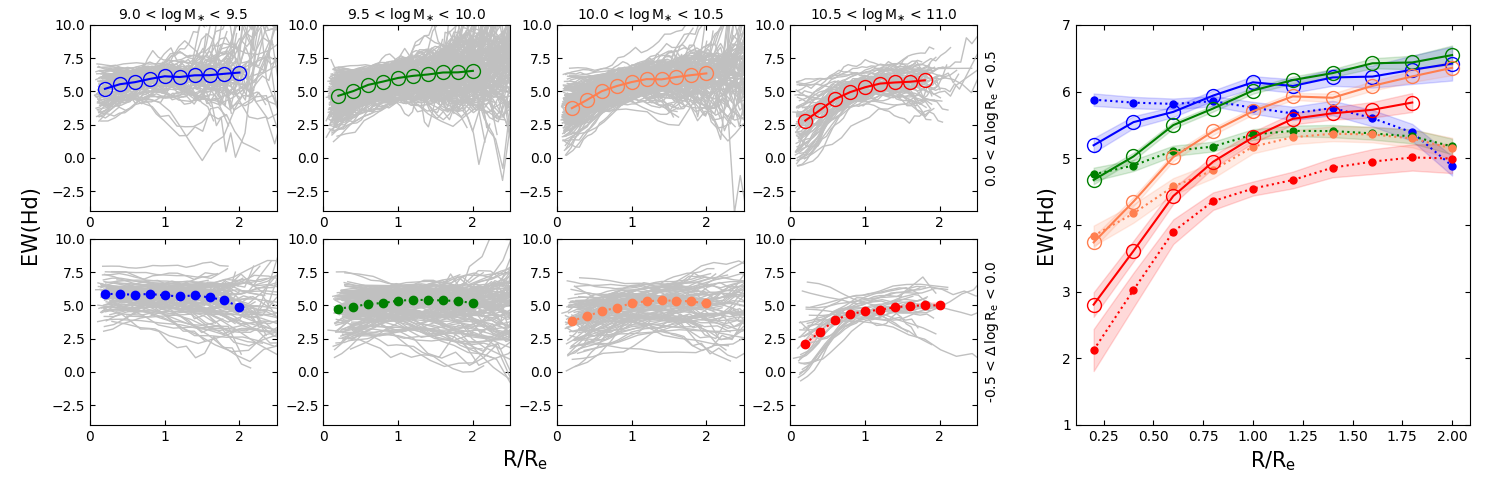}
    \includegraphics[width=0.9\textwidth]{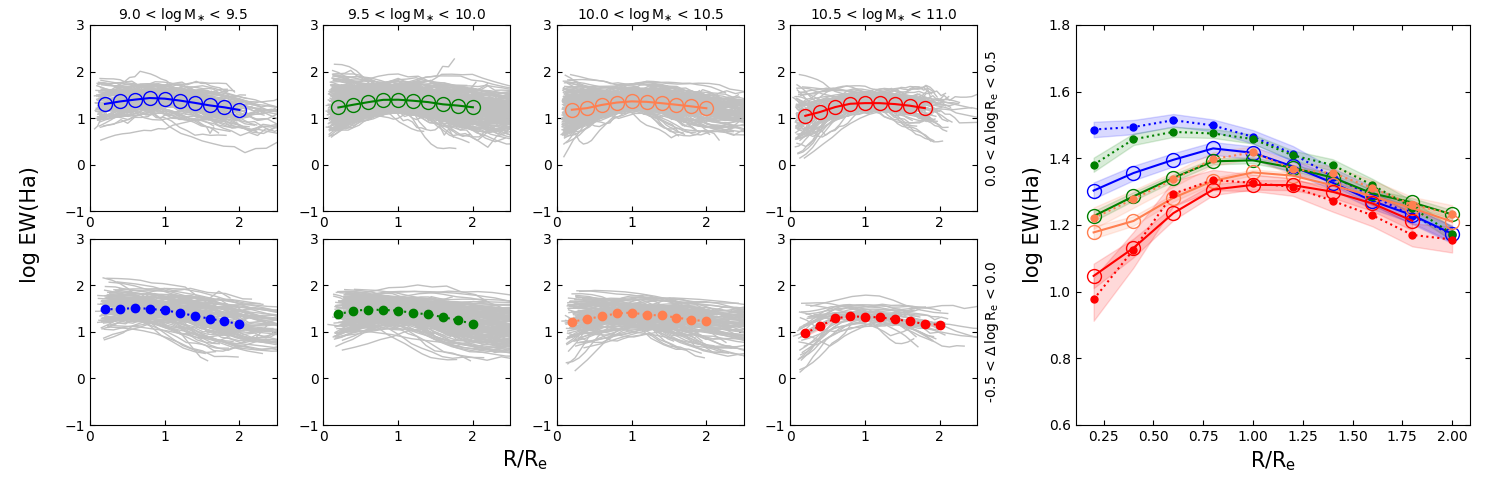}
    \caption{Same as Figure \ref{fig:sigstar_profiles}, but for \Av, \dindex, \ewhd, and \ewha.}
\end{figure*}

\section{PSF effect}
\label{app:psf}

A major concern when comparing the radial profiles of compact and extended galaxies is the difference in their \re/PSF ratios. The beam-smearing effect is more significant for compact galaxies, causing their radial profiles to be systematically biased towards flatter gradients.

Here, we evaluate this effect by matching the datacubes to a similar \re/PSF ratio. Taking the lowest mass bin as an example, the \dlre\ between compact and extended galaxies is $\sim$0.31 dex (a factor of 2). Thus, we aim to convolve the extended galaxies to match a PSF FWHM of 5.0\arcsec, given that the typical PSF FWHM $\sim$2.5\arcsec\ in MaNGA galaxies. For each extended galaxies in the lowest mass bin, we convolve a 2D Gaussian kernel with the 2D image at each wavelength of the IFS data cube. In addition, we bin the raw cube by 2$\times$2 spaxels, generating a new cube of size $N_{x}/2 \times N_{y}/2$, to keep the same spatial sampling as in compact galaxies. We then performed the same analysis on the convolved datacubes. We ran pPXF to fit the continuum, then the stellar mass was estimated by $r-$band luminosity times the light-weighted mass-to-light ($M/L$) ratio. Emission line fluxes were measured after subtracting the continuum, and the SFR was estimated from the dust-corrected H$\alpha$ luminisity as before. The radial profiles were obtained using the same method as in our previous analysis.

\begin{figure*}
	\begin{center}
            \includegraphics[width=0.7\textwidth]{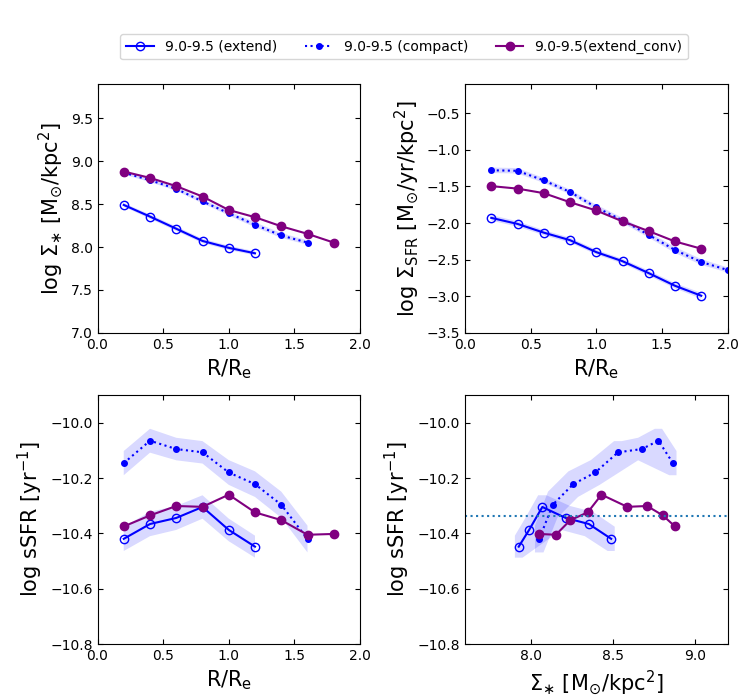}
	\end{center}
	\caption{Upper panels: the \sigstar\ and \sigsfr\ profiles in the convolved extended galaxies (purple lines), compared with the original profiles in compact and extended galaxies (dotted line with filled dots and solid lines with empty circles, respectively) in the lowest mass bin. Lower panels: the sSFR profiles and their relations with \sigstar.}
	\label{fig:psf}    
\end{figure*}

Figure \ref{fig:psf} shows the \sigstar\ and \sigsfr\ profiles in the convolved extended galaxies (purple lines), as well as the original profiles in compact and extended galaxies (dotted line with filled dots and solid lines with empty circles, respectively) in the lowest mass bin. Compared with the original extended galaxies, the convolved extended galaxies have higher \sigstar\ and \sigsfr. By matching the same \re/PSF ratio, compact galaxies and the convolved extended galaxies show similar \sigstar\ profiles. However, the convolved extended galaxies have obvious shallower \sigsfr\ profile than compact galaxies.  The lower panels in Figure \ref{fig:psf} show the sSFR profiles and their relation with \sigstar. The convolved extended galaxies have sSFR profiles similar to the original extended galaxies, but different from compact galaxies. Thus, we conclude that the PSF smearing effect is not sufficient to account for the higher sSFR observed in the inner regions of compact galaxies.

\bibliography{refs}{}
\bibliographystyle{aasjournal}



\end{document}